\documentclass[twocolumn,showpacs,twoside,10pt,prl,superscriptaddress]{revtex4-1}
\usepackage{graphicx}
\usepackage{epsfig}
\usepackage{epsf}
\usepackage{amssymb}
\usepackage{amsmath}
\usepackage{amsthm}
\usepackage{multirow}
\usepackage{mathrsfs}
\usepackage[colorlinks=true,citecolor=blue,urlcolor=blue]{hyperref}
\newcommand\rket[1]{|#1\rangle}
\newcommand\lket[1]{\langle #1|}
\begin{document}

\title{Resonant Quantum Principal Component Analysis}

\author{Zhaokai Li}\altaffiliation{These authors contributed equally to this work.}
\affiliation{Hefei National Laboratory for Physical Sciences at the Microscale and Department of Modern Physics, University of Science and Technology of China, Hefei 230026, China.}
\affiliation{CAS Key Laboratory of Microscale Magnetic Resonance, University of Science and Technology of China, Hefei 230026, China.}
\affiliation{Synergetic Innovation Center of Quantum Information and Quantum Physics, University of Science and Technology of China, Hefei 230026, China.}
\affiliation{Massachusetts Institute of Technology, Cambridge, MA 02139, USA.}

\author{Zihua Chai}\altaffiliation{These authors contributed equally to this work.}
\author{Yuhang Guo}
\author{Wentao Ji}
\author{Mengqi Wang}
\author{Fazhan Shi}
\author{Ya Wang} \email{ywustc@ustc.edu.cn}
\affiliation{Hefei National Laboratory for Physical Sciences at the Microscale and Department of Modern Physics, University of Science and Technology of China, Hefei 230026, China.}
\affiliation{CAS Key Laboratory of Microscale Magnetic Resonance, University of Science and Technology of China, Hefei 230026, China.}
\affiliation{Synergetic Innovation Center of Quantum Information and Quantum Physics, University of Science and Technology of China, Hefei 230026, China.}

\author{Seth Lloyd}
\affiliation{Massachusetts Institute of Technology, Cambridge, MA 02139, USA.}

\author{Jiangfeng Du} \email{djf@ustc.edu.cn}
\affiliation{Hefei National Laboratory for Physical Sciences at the Microscale and Department of Modern Physics, University of Science and Technology of China, Hefei 230026, China.}
\affiliation{CAS Key Laboratory of Microscale Magnetic Resonance, University of Science and Technology of China, Hefei 230026, China.}
\affiliation{Synergetic Innovation Center of Quantum Information and Quantum Physics, University of Science and Technology of China, Hefei 230026, China.}

\begin{abstract}
Principal component analysis has been widely adopted to reduce the dimension of data while preserving the information. The quantum version of PCA (qPCA) can be used to analyze an unknown low-rank density matrix by rapidly revealing the principal components of it, i.e. the eigenvectors of the density matrix with largest eigenvalues.
However, due to the substantial resource requirement, its experimental implementation remains challenging. Here, we develop a resonant analysis algorithm with the minimal resource for ancillary qubits, in which only one frequency scanning probe qubit is required to extract the principal components. In the experiment, we demonstrate the distillation of the first principal component of a 4$\times$4 density matrix, with the efficiency of 86.0\% and fidelity of 0.90. This work shows the speed-up ability of quantum algorithm in dimension reduction of data and thus could be used as part of quantum artificial intelligence algorithms in the future.

\end{abstract}

\maketitle

\section{Introduction}
In many optimization and machine learning applications, principal component analysis (PCA) plays an important role in the process of feature extraction and dimension reduction because of its ability to preserve the information of the data \cite{cPCA_2006,cPCA_2012}.
It is achieved by projecting the data point onto a new low-dimensional basis spanned by the vectors called principal components, which are the eigenvectors of the data set's covariance matrix. To reduce the dimension, one can select only the eigenvectors with large eigenvalues as principal components and discard the ones with eigenvalues below a given threshold. In this way, the variance of the projected data is maximized while the data are mapped into the low-dimensional space. The process of computing the principal components, i.e. the largest eigenvectors of the covariance matrix, involves the diagonalization of a Hermitian matrix and can be speed-up by adopting quantum algorithms. It was shown that a quantum version of principal component analysis (qPCA) \cite{qPCA_2014} is exponentially more efficient than classical methods if the covariance matrix is low-rank and is stored in the form of a quantum state. In combination with recent advances in other linear-algebra-based quantum algorithms such as solving linear systems \cite{HHL_2009,Clader_2013,huang_2019,water_2019}, data analysis \cite{data_2016,wittek_2014}, quantum random accessed memory \cite{qram_2008,park_2019} and learning algorithms \cite{liu_2018,ghosh_2019,havlicek_2019,kapoor_2016,monras_2017,biamonte_2017,svm_exp,wan_2017,beer_2020,benedetti_2017,farhi_2018,perdomo-ortiz_2018,perdomo-ortiz_2019}, this could lead to more applications of quantum machine learning.

The problem of quantum PCA reduces to the question of how to distill the principal components of an unknown low-rank density matrix $\rho=\sum_i \lambda_i \rket{\lambda_i}\lket{\lambda_i}$, where $\langle \lambda_i | \lambda_j \rangle = \delta_{ij}$. If many copies of $\rho$ are given in the quantum form, one can use them to construct the unitary operator $e^{-i\rho t}$ \cite{qPCA_2014,dme_2020}, and then adopts the quantum phase estimation algorithm (PEA) \cite{Nielsen_2000} for the analysis. With the ability of accessing $\rm{log}(\epsilon^{-1})$ ancillary qubits and applying $e^{-i2^k\rho t}$ conditioned on the state of k$-th$ ancillary qubit, PEA can reveal the information of eigenvalues and eigenstates to the accuracy $\epsilon$ within time $O(\rm{poly}(\epsilon^{-1}))$. On an ideal quantum processor, PEA achieves a good level of precision of eigenvalues ($2^{-m}$) given a large number $m$ of the ancillary qubit adopted. However, the demonstration of qPCA remains technically challenging and elusive, due to the high requirements for both the number of qubits and the precision of quantum operations. Furthermore, how far one can reveal the information of the principal components in a coherence-limited physical system is still an open question.

In this work, we propose a resonance-based quantum principal component analysis (RqPCA) algorithm to obtain the principal components by using only one ancillary qubit, with the exponential speed-up being retained. This improvement allows qPCA algorithm to be demonstrated experimentally with current technology, and is scalable to systems containing many quantum bits. We use a prototype hybrid spin system in diamond at ambient conditions, and measure the principal components' eigenvalues with a precision of $2^{-10}$. The length of the quantum circuit scales 
polynomially with the desired accuracy $\epsilon$ of the eigenvalues.
In the experiment, we find that the decoherence of the ancillary qubit becomes the dominant source limiting both the probability of success and the accuracy of the result.
To suppress this effect, the RqPCA algorithm is further developed to combine with the dynamical decoupling strategy, enabling the high-fidelity and high-efficient principal component distillation. The first principal component is distilled from the mixed state $\rho$ with fidelity of 0.90 and the distillation efficiency of 86\%.

\begin{figure*}
\includegraphics[width=1.8\columnwidth]{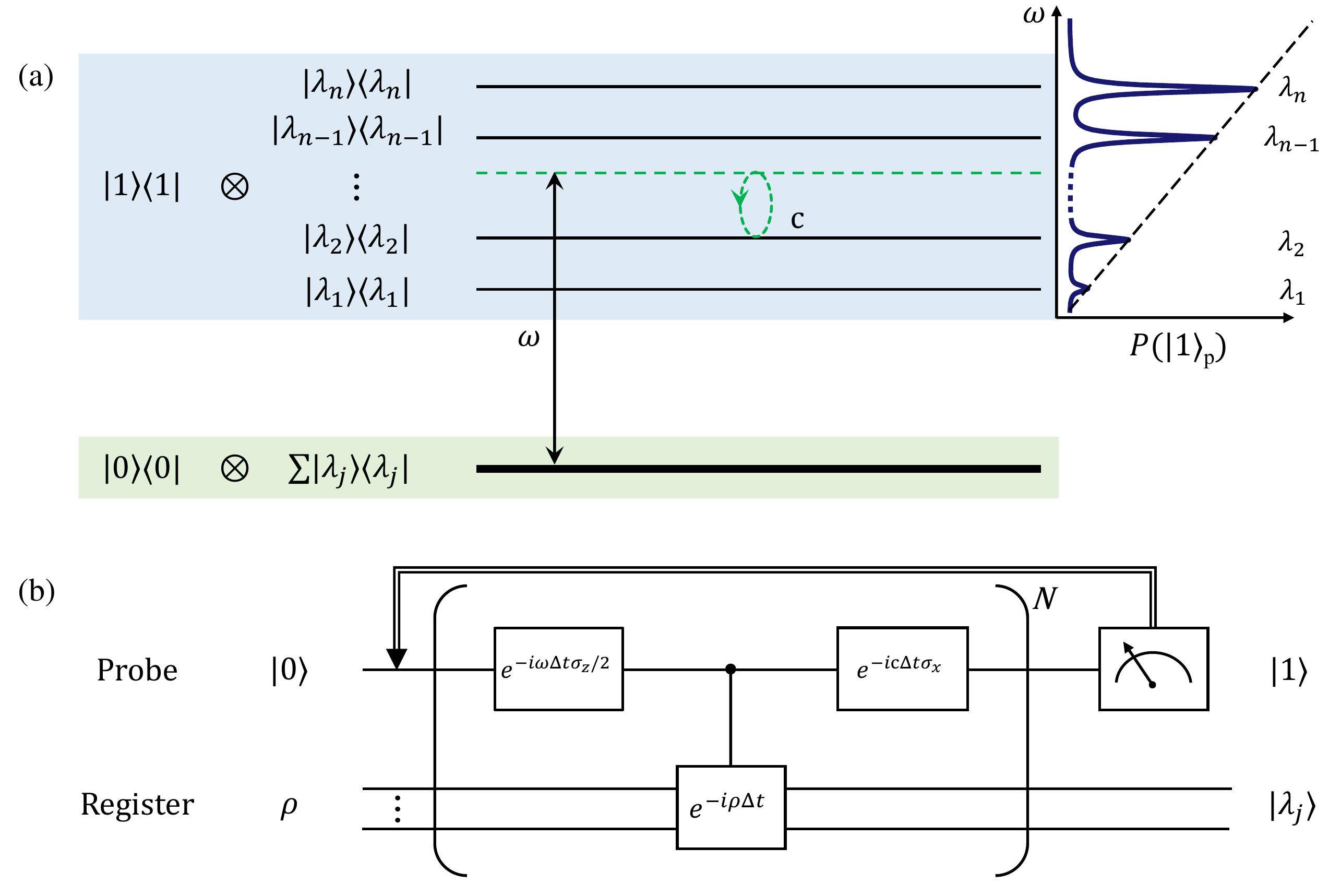}
\caption{Schematics of resonant quantum principal component analysis algorithm. (a) The energy structure of the coupled probe-register system with $H = \rket{1}\lket{1} \otimes \rho$. $\rket{\lambda_i}$ is the $i$-th eigenstate of $\rho$ and $\lambda_i \in [0,1]$ is the corresponding eigenvalue. After introducing the probe qubit's Hamiltonian, the energy of lowest level moves to the green dashed line. Once the scanning frequency $\omega \approx \lambda_i$, the Rabi oscillations of the probe qubit is induced. (b) The quantum circuit of RqPCA using Suzuki-Trotter decomposition ($N\Delta t = \tau$). The projective measurement of the probe qubit in the state $\rket{1}$ indicates success of the algorithm, with principal component being distilled in the register.
}\label{fig1_scheme}
\end{figure*}

\section{Results}
The basic idea of our scheme is illustrated in Fig.~\ref{fig1_scheme}. We start with an ancillary qubit conditionally coupled to the quantum register with the overall Hamiltonian $H = \rket{1}\lket{1} \otimes \rho$. Its time evolution generates exactly the conditional evolution operator $e^{-i\rho t}$ which is the core of qPCA. Then a tunable energy offset was introduced, leading to the Hamiltonian  $H_{int} = \rket{1}\lket{1} \otimes \rho + \frac{\omega}{2} \sigma_z \otimes \rm{I}_n$ where $\sigma_{x,y,z}$ is the Pauli operator of the probe qubit, and $\rm{I}_n$ is the identity matrix which has the same dimension with $\rho$.
The energy spectrum of this system is shown in Fig.~\ref{fig1_scheme}(a), where $\rket{\lambda_n}$ is the eigenstate corresponding to the largest eigenvalue, i.e. the principal component of interest. If a small external field drives the ancillary qubit with strength $c$, the transition between eigenstates $\rket{1}\rket{\lambda_i}$ and $\rket{0}\rket{\lambda_j}$ will be excited when $i=j$ and $|\omega-\lambda_i|$ is small. The ancillary qubit thus probes the transition occurring condition $\omega \approx \lambda_i$ by monitoring its state change.

Given the copies of quantum state of interest $\rho$, we initialize the probe qubit on the state $\rket{0}$ and have the initial state $\rho_{\rm{ini}} = \rket{0}\lket{0} \otimes \rho$. Then the system is evolved under the Hamiltonian

\begin{equation*}\label{Hamiltonian}
\begin{aligned}
&\mathcal{H}_\text{Rq} (\omega) = \frac{w}{2} \sigma_z \otimes {\rm{I}_n} + c \sigma_x \otimes \rm{I}_n + \rket{1} \lket{1} \otimes \rho \\
&= \sum_i \frac{w-\lambda_i}{2} \sigma_z \otimes\rket{\lambda_i}\lket{\lambda_i} + c \sigma_x \otimes {\rm{I}_n} + {\rm{I}_2} \otimes  \sum_i \frac{\lambda_i}{2} \rket{\lambda_i}\lket{\lambda_i}
\end{aligned}
\end{equation*}
for a certain time $\tau$. Once the frequency $\omega$ matches one specific eigenvalue $\lambda_i$ of $\rho$, the probe qubit will flip from $\rket{0}$ to $\rket{1}$ with the probability
\begin{equation*}
  {P_i(\omega)} = \lambda_i D_{i}^2 {\sin ^2}\left( \frac{c \tau }{D_i} \right),i = 1,2,...,n,
\end{equation*}
where $D_i=\sqrt{\frac{(2 c)^2}{(2 c)^2 + (\omega - \lambda_i )^2}}$. The transition probability $P_i(\omega)$ approaches its optimal value $\lambda_i$ in the resonant condition, i.e. ${\omega} - \lambda_i \ll c$ and $\tau \approx \frac{\pi}{2c}$. By scanning the frequency $\omega$ and recording the readout probability being in state $\rket{1}$, one can obtain a typical resonance spectrum, as shown in Fig.~\ref{fig1_scheme}(a), where the position of each resonance peak tells the specific eigenvalue.

\begin{figure*}
\includegraphics[width=1.8\columnwidth]{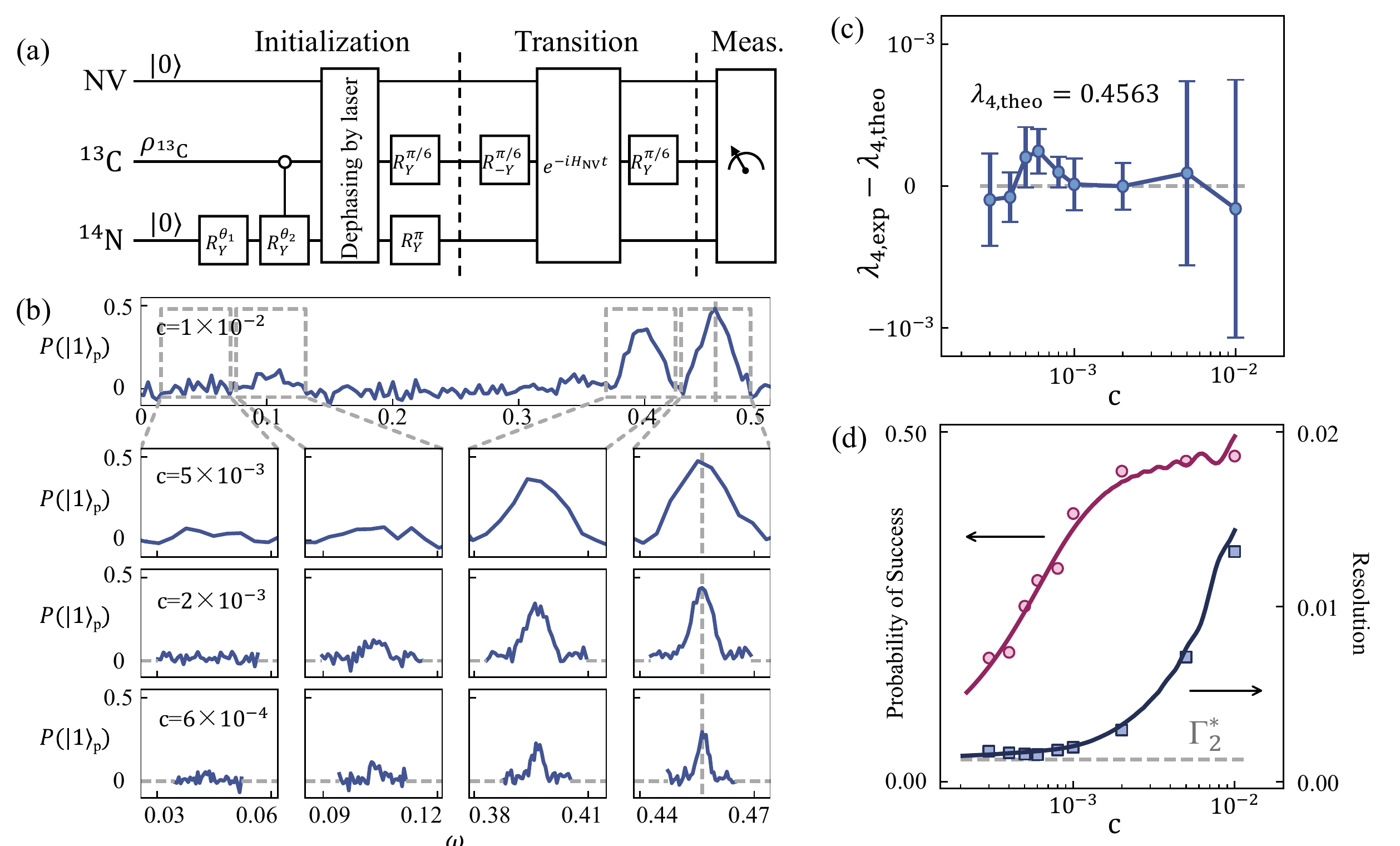}
\caption{The resonant spectra obtained from an adaptive implementation of RqPCA. (a) The schematic circuit of resonant quantum principal component analysis. The electron in NV serves as probe qubit while the nuclear spins are used to store the information of the matrix $\rho$ (see Materials and Methods for details.) (b) The transition spectra obtained through an adaptive implementation of the method. The resonant peaks appear when $\omega$ is close to one of the eigenvalues $\lambda_j$ of $\rho$. (c) The deviations of the experimentally measured eigenvalues of the first principal component $\rket{\lambda_4}$, compared to the theoretical expectations. The experimental eigenvalue and its uncertainty are obtained through a Gaussian spectrum fitting of the transition spectrum in (b). Results of different driving strength $c$ are compared with the same times of circuit repetitions. (d) The probability of success and the resolution of $\lambda_4$, with respect to different driving strength $c$. The dots refer to the experimental results and the solid lines denote the modeled result taking into account the dephasing of the electron spin. The dashed line marks the lower bound of the resolution $\Gamma_2^*=\sqrt{\ln 2} /\left(\pi T_{2e}^{*}f_{\rm{map}}\right)$ due to the dephasing of the probe qubit (see Materials and Methods).
}\label{fig2}
\end{figure*}

After having the probability distribution information, one can quickly locate the eigenstate of interest, e.g. the first principal component $\rket{\lambda_n}$ corresponding to the largest eigenvalue $\lambda_n$. In this case, only the transition from $\rket{0}\rket{\lambda_n}$ to $\rket{1}\rket{\lambda_n}$ is excited, while all other components $\rket{0}\rket{\lambda_j} (j \neq n)$ remain in the subspace of $\rket{0}$. After a projective measurement of the probe qubit, the readout of state $\rket{1}$ indicates that the quantum register is projected into $\rket{\lambda_n}$. If the probe is still in $\rket{0}$, which means no transition was excited, one can return to the algorithm's start and re-run the circuit. The probability of success in a single run equals to $P_n(\omega)$ and is close to $\lambda_n$ in the optimal case, which is the population of the first principal component in the initial state $\rho$.

The evolution of Hamiltonian $\mathcal{H}_\text{Rq}(\omega)$ can be implemented through Suzuki-Trotter decomposition (Fig.~\ref{fig1_scheme}(b)) \cite{trotter}. The controlled operation of $e^{-i\rho\Delta t}$ can be implemented with extra copies of $\rho$ using the method in \cite{qPCA_2014,dme_2020}. In comparison with the conventional qPCA algorithm, RqPCA minimizes the number of ancillary qubits required in quantum phase estimation at the cost of increasing quantum circuit repetitions for the frequency scanning. To further optimize our method, the adaptive implementation is adopted and significantly reduce the repetition times by focusing the area around the eigenvalues of interest. On the other hand, the length of the quantum circuit of RqPCA has the similar scaling property 
as conventional qPCA, with potential complexity advantage benefit from the lower number of qubits. Therefore this method is more applicable to current intermediate-scale quantum computers.

We experimentally demonstrate this algorithm on a nitrogen-vacancy defect (NV) center electron spin associated with the nitrogen nuclear spin (N) and a nearby carbon nuclear spin (C). The electron spin ($\rket{0}:m_S = 0$, $\rket{1}:m_S = +1$) is chosen as the probe qubit, and two nuclear spins ($^{14}$N, $m_I = \{+1,0\} $; $^{13}$C, $m_I = \{+1/2, -1/2\}$) are utilized as the quantum register to store the density matrix $\rho$ for analysis. In this hybrid spin system, electron spins offer fast, versatile and high fidelity readout and control ~\cite{nulea_ssr_2010, elec_ssr_2011, nn_entanglement_measurment_2012, ee_entanglement_2013, ee_remote_entanglement_2013, ee_control_2014, fault-tolerant_control_2015, dynamical_intil_2019, program_control_2019}, and nuclear spins provide additional qubits for the quantum register with long coherence time ~\cite{one_second_2012, e-n_qec_2014,qec_2014,ten_second_storage_2016,ten_qubit_2019}.
The Hamiltonian of the NV-C-N system driven in an external microwave field is described by
\begin{equation*}
\mathcal{H}_{\text{NV}}=\frac{\delta}{2} \sigma_{z}^{e}+\frac{\Omega_{\rm MW}}{2} \sigma_{x}^{e} + \rket{1}_{e}\lket{1} \otimes\left(A_\parallel^{C} I_{z}^{C}+ A_\parallel^{N} I_{z}^{N}\right)
\label{eq:hamil_nv}
\end{equation*}
in the rotating frame of microwave frequency, where $\Omega_{\rm MW}$ is the amplitude of the microwave control field, $A_\parallel^{C} $ and $A_\parallel^{N} $ denote the hyperfine coupling strengths between the electron spin and the two nuclear spins, respectively (see Materials and Methods).

\begin{figure*}
\includegraphics[width=1.8\columnwidth]{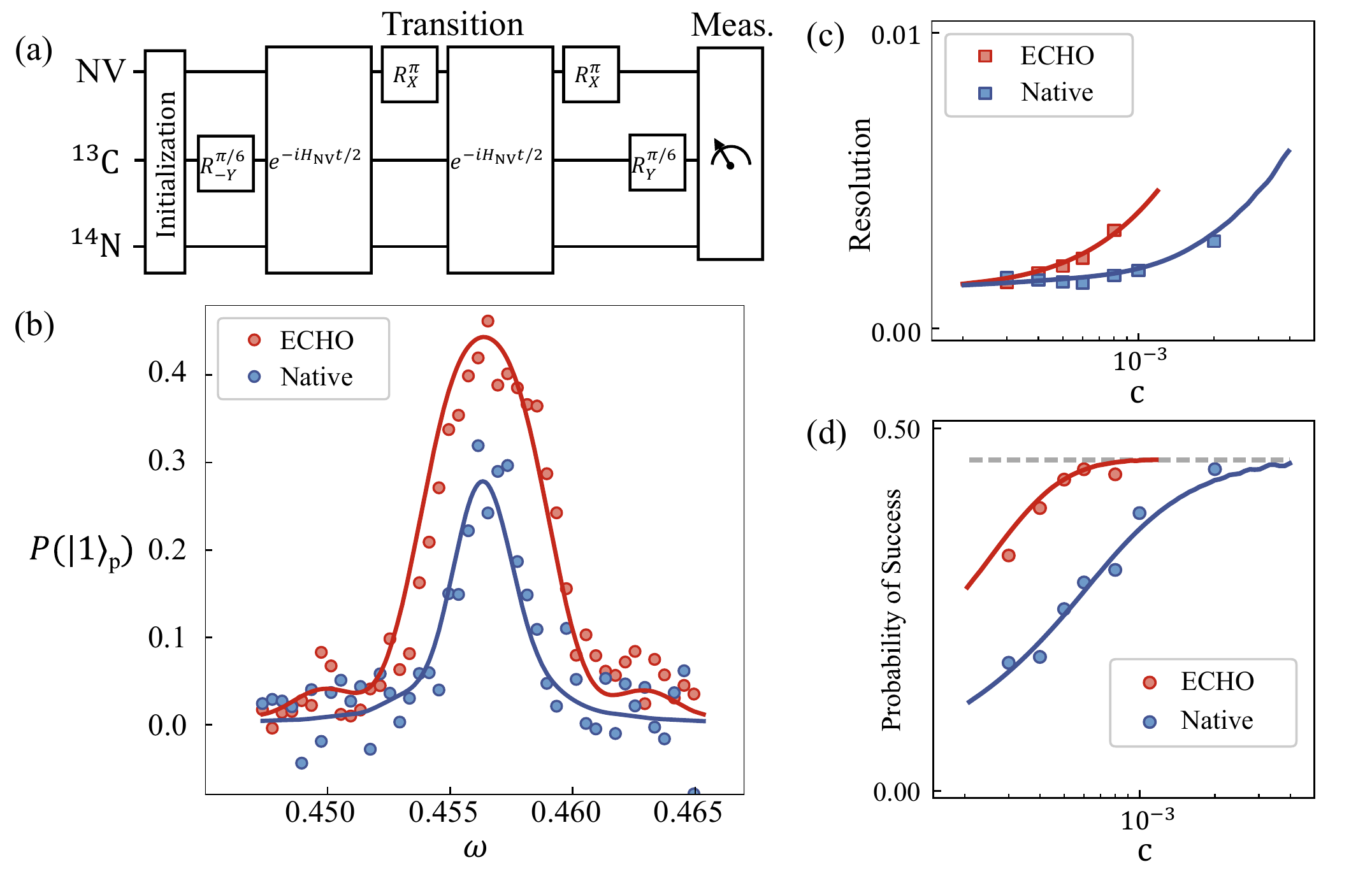}
\caption{RqPCA with ECHO pulses to suppress the decoherence of the probe qubit. (a) Schematic diagram of the ECHO pulses used to suppress the decoherence effect. (see Materials and Methods for details.) (b) The resonant spectral with ECHO (red) and native method (blue) when $c=6 \times 10^{-4}$. (c) and (d) The resolution of the eigenvalue and probability of success with respect to driving strength $c$. The probability of success has a upper bound (dashed horizontal line) that equals to the population of first principal component in $\rho$. Here dots are the experimental results and solid lines are theoretically simulated results with modeled decoherence.
}\label{fig3}
\end{figure*}

The density matrix of interest in this experiment is $\rho = 0.15 \sigma_{z}^{1}+0.09 \sigma_{x}^{1}-0.03 \sigma_{z}^{2}+\rm{I}_4/4$. Since it is a non-product mixed state, a combination of nuclear spin rotation, non-local controlled operation, and a controllable laser-induced dephasing process are adopted to prepare initial state $\rho_{\rm{ini}} = \rket{0}\lket{0} \otimes \rho$. The initialization fidelity reaches value of up to 95$\%$. For a given $\omega$, the corresponding evolution Hamiltonian $\mathcal{H}_{\text{Rq}}(\omega)$ can be constructed from the Hamiltonian of NV system through a local transformation and mapping (see Materials and Methods). Fig.~\ref{fig2}(a) shows the experimental diagram proceeding in three steps. The evolution time is setting as $ \tau = \pi / 2c$ so that the transition probability is optimized. Finally, the electron spin state is optically readout to get the transition possibility for different $\omega$, from which the eigenvalues are obtained directly.

Fig.~\ref{fig2}(b) shows the transition spectrum obtained through an adaptive implementation of the method. A considerable large driving strength is firstly applied to estimate the eigenvalues in a broad frequency range quickly. After this, the driving strength and the scanning range are tuned adaptively according to the previous step's resonance peak
information. In four adaptive steps, the spectral line-width (resolution) approaches to a lower bound while the number of sampling points keeps low. In combination with the adaptive method, RqPCA can thus significantly reduce the frequency scanning repetition times. However, the enhancement of the resolution is at the cost of longer experiment length and therefore losing the probability of success due to the decoherence effect in realistic experiments  (Fig.~\ref{fig2}(d)). Thus, the uncertainty of the peak position decreases initially with the reduction of spectral line-width but lately increases due to the low success probability (Fig.~\ref{fig2}(c)). In the experiment, the observed peak position has a maximal deviation of $3 \times 10^{-4}$ to the theoretical eigenvalue caused due to the external magnetic field instability (see Materials and Methods). The resulting eigenvalue precision is $2^{-10}$, equivalent to perfect PEA implementation with ten ancillary qubits in the conventional qPCA algorithm.


\begin{figure*}
\includegraphics[width=1.8\columnwidth]{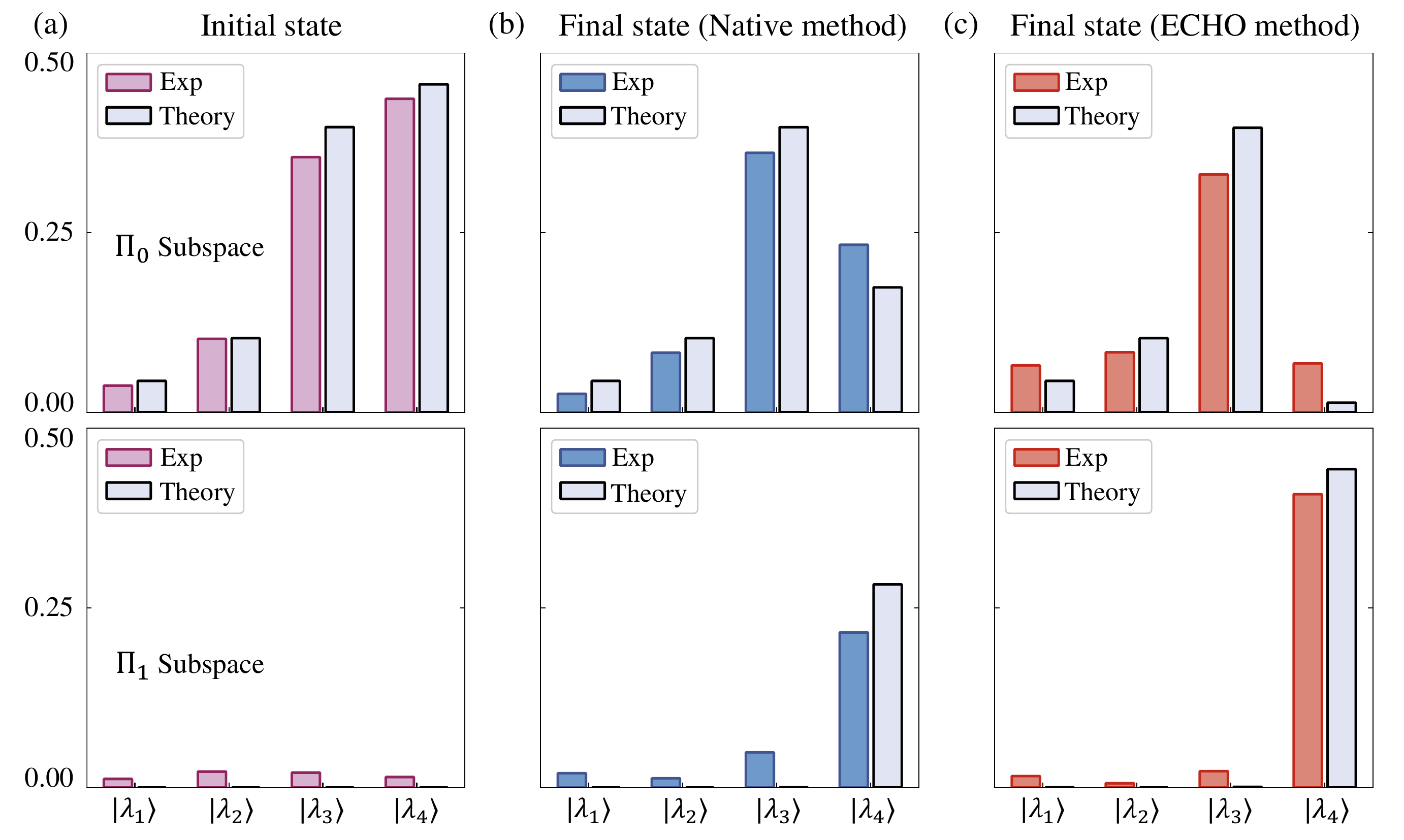}
\caption{The principal component distillation. (a) The initial state before RqPCA obtained by state tomography. (b) and (c) The states after applying the native and ECHO method of RqPCA, respectively, with the same parameters: $c=6 \times 10^{-4}$, and $\omega = 0.4566$ determined from Fig.~\ref{fig3}(b). Here, the subspace $\prod_{k}$ is spanned by the eigenbasis of $\rho$ ($\rket{\lambda_{1,2,3,4}}$) while the probe (electron spin) is in the state $\rket{k} (k=0,1)$. The bars denote the populations of the basis states $\rket{\lambda_{1,2,3,4}}$ in the experimentally measured states.
}\label{fig4}
\end{figure*}

To better understand the driving strength dependent effect mentioned above, we consider the dephasing of electron spin using independently measured decoherence quantities. We find good agreement between the modeled results (solid lines in Fig.~\ref{fig2}(d)) and the experimental results (dots) of both the uncertainties of the eigenvalue and success probabilities. The success probability reduces since the electron spin loses its coherence during its flip to $\rket{1}$, due to the dephasing process. Thus the resolution of the eigenvalue is limited by the dephasing rate of the electron spin. To suppress the dephasing and increase the success probability, we extend the current scheme to an approach that naturally combines with the dynamical decoupling technique \cite{dd_2009}. The single evolution period is split into two half periods with the first $\pi$ pulse applied at the middle, and the second $\pi$ pulse used at the end (Fig.~\ref{fig3}(a)). The resonant spectral with this modification (red, denoted as ECHO method below) and native method (blue) when $c=6 \times 10^{-4}$ are shown in Fig.~\ref{fig3}(b). While using the ECHO method, the probability of success is increased significantly, leading to the higher efficiency of principal component distillation in later steps. Fig.~\ref{fig3}(c) shows how the likelihood of success and resolution of the eigenvalue varies with different $c$. When $c$ is small, the possibility of success adopting the ECHO method is always much higher than the one in the native way. In contrast, the resolution of both methods approaches to a lower bound which comes from the dephasing of the electron spin.

To reveal the first principal component $\rket{\lambda_4}$, the resonance frequency $\omega$ is tuned to be the largest eigenvalue $\lambda_4$ measured in previous steps, so only the first principal component will be on resonant. After applying the circuit of RqPCA, if the probe qubit is measured as $\rket{1}$, the other qubits will collapse to the first principal component of $\rho$, i.e. $\rket{\lambda_4}$. To better understand how well the principal component is distilled, we measured both the state before and after the RqPCA using state tomography (see Materials and Methods). The first column of Fig.~\ref{fig4} shows the measured initial state represented in terms of four eigenstates of $\rho$. A mixture of four eigenstates is observed in the subspace of probe being in $\rket{0}$, while almost no population in the subspace of $\rket{1}$. The same measurements are performed after RqPCA, and the results using native and ECHO method are shown in the second and third columns, respectively. After running the circuit, the state in the subspace of $\rket{1}$ is close to the principal component $\rket{\lambda_4}$, when the state in the subspace of $\rket{0}$ remains a mixture of different eigenstates.
It is noted that although both methods (ECHO and native) can distill the principal component $\rket{\lambda_4}$ from the background of the mixed eigenstates in $\rho$, the efficiencies of distillation are different. While adopting the ECHO method, 86.0\% of the population of $\rket{\lambda_{4}}$ transits to the subspace of $\rket{1}$, which indicates a high efficiency of principal component distillation. At the same time, very few populations of other components ($\rket{\lambda_{1,2,3}}$) appears in the subspace of $\rket{1}$, leading to a distillation fidelity of 0.90. In the case of the native method, both the efficiency (48.1\%), and the fidelity (0.73) are much lower due to the electron spin's dephasing effect.

The resonance-based principal component analysis algorithm in this work optimizes the resource requirement for ancillary qubits in the phase estimation step. It can be easily applied to resolve other eigenstate finding problems such as molecular energy simulation in quantum chemistry ~\cite{quantum_chemistry_2005}. If combined with a faithful demonstration of density matrix exponentiation, our method could serve as an essential part of a variety of quantum machine learning implementations. Furthermore, the ability to combine with decoherence suppressed technique, makes the method applicable to the intermediate-scale noisy quantum computers nowadays.

\section{Materials and Methods}

\textbf{Experimental setup and spin system.} The experiment was carried out on a home-built confocal microscope at ambient conditions. A continuous wave laser at 532 nm is used for optical pumping and readout of the NV spin, and is gated with two acoustic-optic modulators (AOM). The laser beam was focused by an oil objective, while the fluorescence signal was collected by the same objective. An active temperature control to within 5 mK was used to increase the magnetic field stability. The microwave signal used to control the electron spin was generated by an arbitrary waveform generator(AWG, crs1w000b, CIQTEK) in combination with a microwave generator through the I/Q modulation. The radio-frequency signal used to control the nuclear spins was also generated by the same AWG.

We used a NV center containing one coupled $^{13}$C nuclear spin ($A_\parallel^{C} \simeq 12.8\ \mathrm{MHz}$) and an intrinsic $^{14}$N nuclear spin ($A_\parallel^{N} \simeq -2.16\ \mathrm{MHz}$) in a [100]-oriented diamond. To improve the photon collection efficiency, a solid immersion lens was fabricated on the NV centers. An external magnetic field of 380 Gauss was applied to remove the degeneracy between the electronic states $m_S = +1$ and $m_S = -1$.
The dephasing time of the whole spin system are measured as $T_{2,\rm e}^* \sim $ 5.8 $\mu$s, $T_{2,\rm C}^* \sim $ 2.0 ms  and $T_{2,\rm N}^* \sim $ 5.0 ms. In comparison to the evolution time of the whole circuit, only the dephasing of electron spin is dominant for the effect of decoherence.

\textbf{Hamiltonian mapping.} By applying a rotation of angle $\theta^{'}$ on the $^{13}$C nuclear spin, the Hamiltonian $\mathcal{H}_{\text{NV}}$ is transformed to:
\begin{equation*}
\mathcal{H}=\frac{\delta}{2} \sigma_{z}^{e} + \frac{\Omega_{\rm MW}}{2} \sigma_{x}^e +|1\rangle_e\langle{1}|\otimes \left(\alpha I_{z}^{C} + \beta I_{x}^{C} + A_{z}^{N} I_{z}^{N}\right)
\label{eq:hamil_nv_2}
\end{equation*}
where $\alpha=A_z^C \cos{\theta^{'}}, \beta=A_z^C \sin{\theta^{'}}$. By setting $\delta=f_{\rm{map}}(\omega-1/4)$ and $\Omega_{m w}= 2f_{\rm{map}} c$ with a mapping factor $f_{\rm{map}}=2\pi\times36.25$ MHz, the Hamiltonian of NV is mapped to the evolution Hamiltonian $\mathcal{H}_{\text{Rq}}(\omega)$ with a difference of constant term that can be neglected.

\textbf{State preparation.} Due to dynamical optical pumping~\cite{dnp_2009}, the spin system starts from an initial state $\rho_0=|0\rangle_e\langle0|\otimes\rho_{\rm C}\otimes|0\rangle_{\rm N}\langle0|$, with $\rho_{\rm C}=\left(\begin{array}{cc}
0.85 & 0 \\
0 & 0.15
\end{array}\right)$. The spin rotation $\mathrm{R}_{Y}(\theta_1) : = e^{-i{\sigma_y} \theta_1/2}$ on the nitrogen nuclear spin and a non-local $\mathrm{C}^{\rm C} \mathrm{ROT}^{\rm N}(\theta_2)$ gate then distribute populations among nuclear spin state subspace. Here $\mathrm{C}^{\rm C} \mathrm{ROT}^{\rm N}(\theta_2)$ denotes the spin rotation of nitrogen nuclear spin $\mathrm{R}_{\rm Y}^{\rm N}(\theta_2)$ conditioned on the carbon nuclear spin being in state $\rket{0}$. This is realized by a conditional phase gate on the electron spin, in combination with local controls \cite{e-n_qec_2014}. As can be seen in Fig.~\ref{figs1}, the electron spin transition spectrum associated with nitrogen state $\rket{1}_{\rm N}$ changed to the one associated with nitrogen state $\rket{0}_{\rm N}$ after the $\mathrm{C}^{\rm C} \mathrm{ROT}^{\rm N}(\pi))$ gate, while it did not change when the carbon nuclear spin state is in the state $\rket{0}_{\rm C}$.  The resulting state is:
\begin{equation*}
0.85|0\rangle_{\rm C}\langle 0|\otimes\left(\begin{array}{cc}
\alpha_{1}^{2} & \alpha_{1} \beta_{1} \\
\alpha_{1} \beta_{1} & \beta_{1}^{2}
\end{array}\right)
+0.15| 1\rangle_{\rm C}\langle 1| \otimes\left(\begin{array}{cc}
\alpha_{2}^{2} & \alpha_{2} \beta_{2} \\
\alpha_{2} \beta_{2} & \beta_{2}^{2}
\end{array}\right),
\end{equation*}
where $\alpha_1=\cos(\frac{\theta_1+\theta_2}{2})$, $\beta_1=\sin(\frac{\theta_1+\theta_2}{2})$, $\alpha_2=\cos(\frac{\theta_1}{2})$ and $\beta_2=\sin(\frac{\theta_1}{2})$. After this, a laser-induced dephasing process is introduced to eliminate the off-diagonal matrix terms. As shown in Fig.~\ref{figs1}, at a laser power of 190 $\mu$w, a fast dephasing time ($T_{2,\rm laser}^*\sim 0.6 \mu$s) of nitrogen nuclear spin was observed, accomplished by a repolarization to the state $|0\rangle_{\rm N}\langle0|$ with a decay rate of $T_{1,\rm laser}\sim 2.1 \mu$s. Based on these results, we choose a laser pulse length of 1.4 $\mu$s to completely dephase the coherence, and $\theta_1 = 0.58 \pi$ and $\theta_2 = 0.31 \pi$ to account for the finite repolarization. The quantum state turns into:

\begin{equation*}
0.85|0\rangle_{\rm C}\langle 0|\otimes\left(\begin{array}{cc}
0.53 & 0 \\
0 & 0.47
\end{array}\right)
+0.15| 1\rangle_{\rm C}\langle 1| \otimes\left(\begin{array}{cc}
0.70 & 0 \\
0 & 0.30
\end{array}\right).
\end{equation*}
Finally a single-qubit rotation $\mathrm{R}_{Y}^{\rm C}(\frac{\pi}{6})$ on the carbon nuclear spin, and a $\mathrm{R}_{Y}^{\rm N}(\pi)$ pulse on the nitrogen nuclear spin were applied, the system was prepared into the state $\rho$ with fidelity $95\%$. Fig.~\ref{figs2} shows the density matrix obtained by the state tomography.

\textbf{Error analysis.} To understand the deviation between experimental and theoretical eigenvalues, we notice that the external magnetic field slowly drifts during the experiments. Suppose $B_0$ changes $\delta{B}$ during the experiments, then the Hamiltonian $\mathcal{H}$ becomes:
\begin{small}
\begin{equation*}
\mathcal{H}_{\rm{drift}}=\frac{(\delta-\gamma_e{\delta_{B}})}{2} \sigma_{z}^{e} + \frac{\Omega_{\rm MW}}{2} \sigma_{x}^e +|1\rangle_e\langle{1}|\otimes \left(\alpha I_{z}^{C} + \beta I_{x}^{C} + A_{z}^{N} I_{z}^{N}\right).
\end{equation*}
\end{small}
According to the linear mapping between $\mathcal{H}$ and $\mathcal{H}_{\text{Rq}}(\omega)$, the effective $\omega$ in $\mathcal{H}_{Rq}(\omega)$ changes to:
\begin{equation*}
\omega_{\rm{eff}}=\omega-\delta \omega=\omega-\gamma_{e} \delta_B / {f_{\rm{map}}}
\end{equation*}
The change of external magnetic field $\delta_B$ thus causes the deviation of $\omega$ in the experiments. By estimating the instability of magnetic field via recording the electron spin's resonance frequency in 14 hours, we find $\delta_B$ satisfies a Gaussian distribution and the standard deviation is 12.4 kHz, which corresponds to $\delta \omega=3.4\times 10^{-4}$, consistent with the eigenvalue inaccuracy observed in the experiment.

\textbf{Dynamical decoupling combined component analysis.} Here we consider the effect of dephasing of the electron spin while using a general sequence in the form of $(\frac{\tau}{2M}-\pi-\frac{\tau}{2M}-\pi)^M$, where $\tau=\pi/2c$. The Native and ECHO method corresponds to the case of $M$ = 0 and $M$ = 1 respectively.  The dephasing is considered by adding an additional term $\delta_f \sigma_z/2$ into the total Hamiltonian $\mathcal{H}_\text{Rq} (\omega)$, with $\delta_f$ satisfying a Gaussian distribution with the standard deviation determined by $T_{2,e}^{*}$. For the Native case, this model returns to a pulsed spin resonance with a spectral width $\sqrt{\ln 2} /\left(\pi T_{2,e}^*\right)$ ~\cite{odmr_2011}.

A more general consideration of the dephasing is performed through the numerical simulation. Fig.~\ref{figs3} shows the spectra and the extracted features of resolution and amplitude, from $M=0$ to $M=8$. One can see that when the driving strength $c$ is very weak, increasing the number $2M$ of $\pi$ pulses can continuously improve peak's amplitude to the upper bound, with almost no broadening of peak-width.

\textbf{Efficiency of RqPCA.} Compared to the qPCA using phase estimation algorithm, RqPCA minimizes the number of ancillary qubits needed at the cost of increasing quantum circuit repetitions for the frequency scanning. Both methods need the ability of applying the conditional evolution operator $e^{-i\rho t}$ with the help of multiple copies of $\rho$, and require the evolution time $t=O(\epsilon^{-1})$ to achieve the accuracy $\epsilon$ of the eigenvalues. Following the smaller scale of the quantum circuit in RqPCA, the number and complexity of multi-qubit quantum gates are also reduced. Therefore, the cost of single experiment in our RqPCA method is much lower, enabling the high-fidelity and high-efficient experimental implementation. Although the RqPCA requires more quantum circuit repetitions to obtain the resonant spectrum, the adaptive implementation used in this work can significantly reduce the repetition times by only focusing the area around the eigenvalues of interest. In the worst case, the repetitions time needed to obtained the resonant spectrum scales as $O(\epsilon^{-1})$, which also scale polynomially with the desired accuracy of the eigenvalues.

\bibliography{references_PCA}
\bibliographystyle{apsrev4-1}

\textbf{Acknowledgements.}

This work is supported by the National Key R$\&$D Program of China (Grant No.\ 2018YFA0306600, 2017YFA0305000), the National Natural Science Foundation of China (Grants No.\ 11775209, 81788101, 11761131011, 11575173), the CAS (Grants No. GJJSTD20170001, No. QYZDY-SSW-SLH004) and Anhui Initiative in Quantum Information Technologies (Grant No. AHY050000).  S.\ Lloyd was supported by DOE, NSF, AFOSR and ARO. The authuors thank Hefeng Wang for helpful discussions and suggestions. Z.\ Li thanks the China Scholarship Council for the support.

\renewcommand{\thefigure}{A\arabic{figure}}

\setcounter{figure}{0}

\begin{figure*}[p]
\includegraphics[width=1.95\columnwidth]{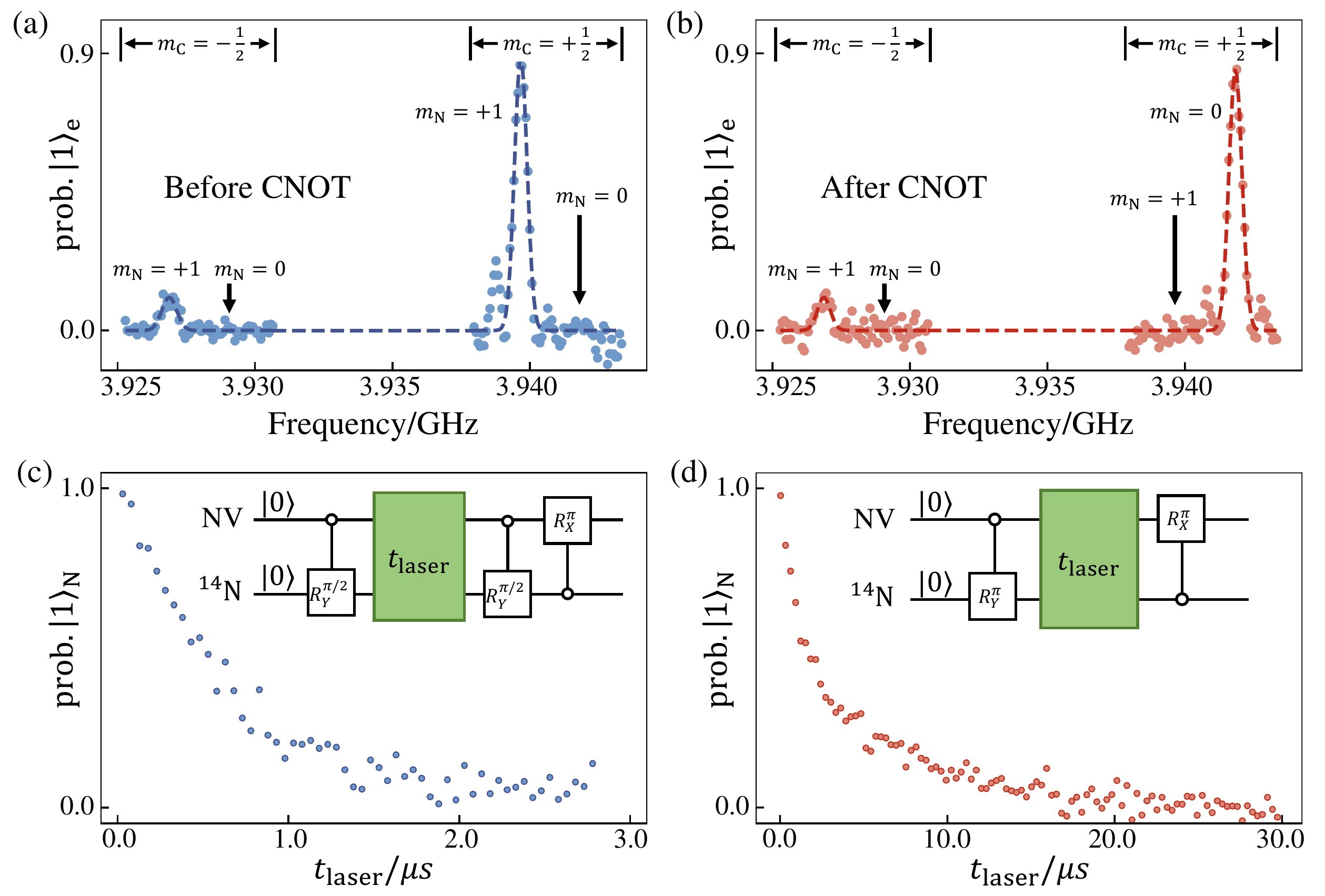}
\caption{Initial state preparation. (a) and (b) The pulsed ODMR spectrum of laser initialized state $\rho_0$ and the state after applying $\mathrm{C}^{\rm C} \mathrm{NOT}^{\rm N}$ gate. (c) The decay of $^{14}$N nuclear spin's coherence under the green laser illumination. The pulse sequence is shown in the inset, including the preparation of the coherent state $(\rket{0}_{\rm N}+\rket{1}_{\rm N})/\sqrt{2}$, the laser-induced dephaing and map to the electron spin for optical readout. (d) The repolarization of $^{14}$N nuclear spin from the state $\rket{1}_{\rm N}$ to the state $\rket{0}_{\rm N}$ under laser illumination. }\label{figs1}
\end{figure*}

\begin{figure*}[p]
\includegraphics[width=1.95\columnwidth]{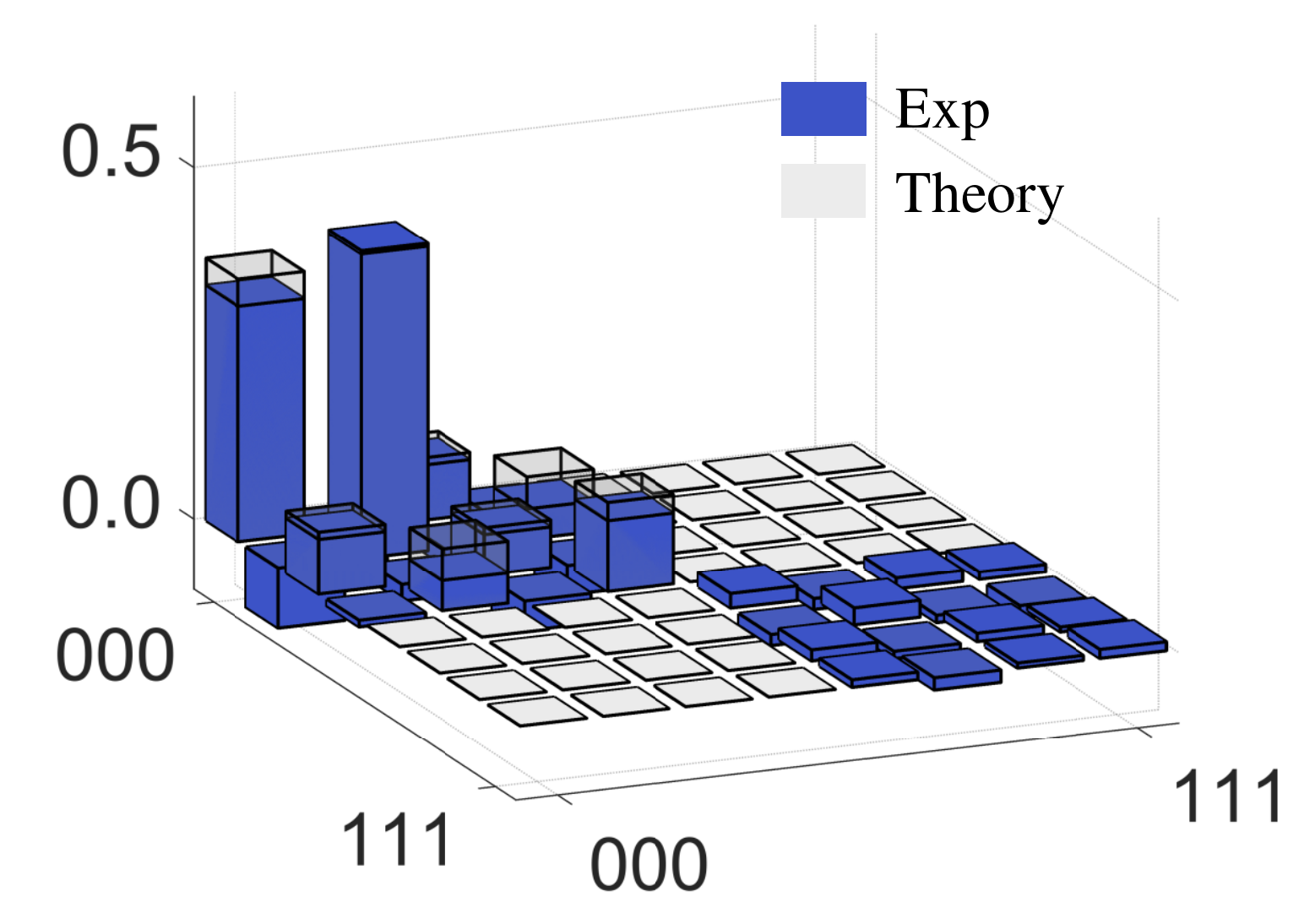}
\caption{{Experimental measured results of the initial state $\rho_{\rm{ini}}$. Only the diagonal blocks relevant for our analysis are shown here.}}\label{figs2}
\end{figure*}

\begin{figure*}[p]
\includegraphics[width=1.95\columnwidth]{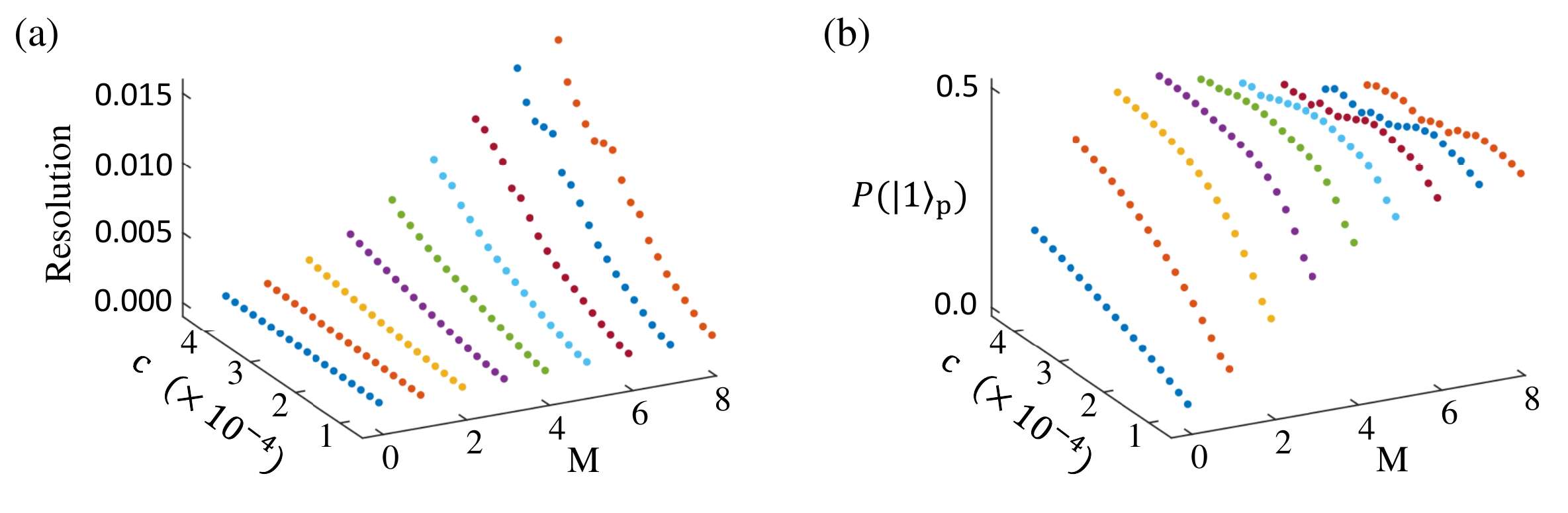}
\caption{Numerical simulation for dynamical decoupling combined principal component analysis.  (a) Resolution obtained from resonant spectra for different driving strength $c$ and the order of echo $M$.  (b) The probability of success obtained from resonant spectra for different $c$ and $M$.}\label{figs3}
\end{figure*}

\end{document}